\documentclass[%
 reprint,
superscriptaddress,
showpacs,
 amsmath,amssymb,
 aps,
floatfix,
]{revtex4-1}

\usepackage{dcolumn}
\usepackage{bm}

\usepackage{graphicx} 
\usepackage{color}

\usepackage{ulem} 

\usepackage[dvipdfmx,
colorlinks=true, citecolor=blue, urlcolor=blue, linkcolor=blue,
setpagesize=false, bookmarks=false
]{hyperref}

\begin{document}
\title{Characterization of photoexcited states in the half-filled one-dimensional extended Hubbard model assisted by machine learning}
\author{Kazuya Shinjo}
\affiliation{Department of Applied Physics, Tokyo University of Science, Tokyo 125-8585, Japan}
\author{Shigetoshi Sota}
\affiliation{Computational Materials Science Research Team,
RIKEN Center for Computational Science (R-CCS), Kobe, Hyogo 650-0047, Japan}
\author{Seiji Yunoki}
\affiliation{Computational Materials Science Research Team,
RIKEN Center for Computational Science (R-CCS), Kobe, Hyogo 650-0047, Japan}
\affiliation{Computational Condensed Matter Physics Laboratory,
RIKEN Cluster for Pioneering Research (CPR), Wako, Saitama 351-0198, Japan}
\affiliation{Computational Quantum Matter Research Team,
RIKEN Center for Emergent Matter Science (CEMS), Wako, Saitama 351-0198, Japan}
\author{Takami Tohyama}
\affiliation{Department of Applied Physics, Tokyo University of Science, Tokyo 125-8585, Japan}

\date{\today}
             

\begin{abstract}
Photoinduced nonequilibrium states can provide new insight into dynamical properties of strongly correlated electron systems.
One of the typical and extensively studied systems is the half-filled one-dimensional extended Hubbard model (1DEHM).
Here, we propose that the supervised machine learning (ML) can provide useful information for characterizing photoexcited states in 1DEHM.
Using entanglement spectra as a training dataset, we construct neural network.
Judging from the trained network, we find that bond-spin-density wave (BSDW) order can be enhanced in photoexcited states if the frequency of a driving pulse nearly resonates with gap.
We separately calculate the time evolution of local and non-local order parameters and confirm that the correlation functions of BSDW are actually enhanced by photoexcitation as predicted by ML.
The successful prediction of BSDW demonstrates the advantage of ML to assist characterizing photoexcited quantum states.
\end{abstract}
\maketitle

%
\section{introduction}\label{sec_1}
Nonequilibrium processes in strongly correlated electron systems can provide new insight into dynamical properties of these systems.
One such example is nonequilibrium induced phase transitions. 
As the system is driven away from the equilibrium, under certain conditions, a ``crossover'' from one state to another (metastable) state may occur, which gives rise to not only a thermally accessible state but also leads to a hidden state of matter. 

Not only insulator-to-metal transitions~\cite{Taguchi2000, Oka2003, Oka2005, Okamoto2007, Takahashi2008, Oka2008, Al-Hassanieh2008} but also an enhancement of charge and bond orders~\cite{Lu2012, Shao2018} as well as superconducting correlations~\cite{Kaneko2019,Kaneko2019_2,Ejima2020} have been suggested as an indication of a nonequilibrium induced phase transition in a one-dimensional (1D) Mott insulator. 
However, the characterization of the emergent nonequilibrium state has not yet been well-established.
One of the possible strategies for such a characterization is the use of machine learning (ML), which can detect hidden features of states without explicitly defining order parameters~\cite{Carrasquilla2017, Broecker2017, Chng2017, Schindler2017, Venderley2018, Ohtsuki2016, Zhang2017, Zhang2017_2, Ohtsuki2017, Broecker2017_2, Nieuwenburg2017, Deng2017, Yoshioka2018, Suchsland2018, Bukov2018, Mehta2018, Rem2018, Bohrdt2018}.

In this paper, we investigate photoexcited states of the half-filled 1D extended Hubbard model (1DEHM).
We propose that supervised ML can provide new insight into characterization of photoexcited quantum states.
Here, entanglement spectrum (ES) of the typical ground states~\cite{Voit1992, vanDongen1994, Nakamura1999, Sengupta2002, Jeckelmann2002, Tsuchiizu2002, Giamarchi, Zhang2004, Sandvik2004, Tsuchiizu2004, Ejima2007, Kumar2009, Kumar2010, Ejima2016, Hafez-Torbati2017} of the half-filled 1DEHM such as the Mott-insulating (MI), charge-density-wave (CDW), and bond-charge-density-wave (BCDW) states are used as a training dataset.
In addition, we also use ES of the bond-spin-density-wave (BSDW) state~\cite{Japaridze1995, Japaridze1997, Japaridze1999, Aligia2000, DiLiberto2014, Nakamura2000, Dobry2010}, which is stabilized by introducing a correlated-hopping interaction to the 1DEHM.
This is because such a correlated-hopping interaction is expected to appear in a Floquet effective model of the periodically driven Hubbard model.
The trained neural network finds that BSDW order can be induced in photoexcited states, although the BSDW state is not a ground state of the 1DEHM.
In order to examine the predictions obtained by ML, we separately calculate the time evolution of local order parameters (LOPs) of BCDW, BSDW, and CDW and the parity and string nonlocal order parameters (NLOPs).
We find that order parameters related to BSDW are enhanced by photoexcitation, which is also supported by the Floquet theory.
This result is in accordance with the prediction obtained by ML.
Thus ML can successfully characterize photoexcited states of the half-filled 1DEHM.

This paper is organized as follows. 
We introduce the Hamiltonian of 1DEHM in Sec.~\ref{sec_2}. 
Especially, we note that the correlated-hopping interaction determines whether BCDW or BSDW is realized in 1DEHM. 
By deriving the Floquet effective Hamiltonian, we show that the correlated-hopping interaction is controlled by periodic field, and thus there is a possibility that BSDW order, which is not present in the phase diagram of 1DEHM, can be enhanced.
We demonstrate in Sec.~\ref{sec_3} that photoinduced enhancements of not only CDW order suggested in Ref.~\cite{Lu2012} but also BSDW order are captured by ML.
In Sec.~\ref{sec_4}, the prediction by ML being a clue, we further investigate photoexcited states of 1DEHM by explicitly calculating LOPs and NLOPs and discuss microscopic origin of the emergence of BSDW.
Finally, a summary is given in Sec.~\ref{sec_5}.

\section{Model}\label{sec_2}
The 1DEHM is defined as 
\begin{align}
\mathcal{H}=&-t_\mathrm{h}\sum_{i,\sigma} B_{i,i+1,\sigma}
+ U\sum_i n_{i,\uparrow}n_{i,\downarrow} + V\sum_i n_{i}n_{i+1}, 
\end{align}
where $B_{i,j,\sigma}=c_{i,\sigma}^\dag c_{j,\sigma}+c_{j,\sigma}^\dag c_{i,\sigma}$, 
$c^\dagger_{i,\sigma}$ is the creation operator of an electron with spin $\sigma (= \uparrow, \downarrow)$ at site $i$, and $n_i=\sum_\sigma n_{i,\sigma}$ with $n_{i,\sigma}=c^\dagger_{i,\sigma}c_{i,\sigma}$. 
We consider the parameter region with $U>0$ and $V>0$ at half filling, taking the nearest-neighbor hopping $t_\mathrm{h}$ to be the unit of energy ($t_\mathrm{h}=1$).
The ground-state phase diagram of the model is well established~\cite{Voit1992, vanDongen1994, Nakamura1999, Sengupta2002, Jeckelmann2002, Tsuchiizu2002, Giamarchi, Zhang2004, Sandvik2004, Tsuchiizu2004, Ejima2007, Kumar2009, Kumar2010, Ejima2016, Hafez-Torbati2017}.
BCDW is an intermediate phase between MI and CDW phases up to a critical value $U=9.25$~\cite{Ejima2007}.
In addition, we can have the ground state with bond order, i.e., BSDW (BCDW) order, if we introduce the correlated-hopping term 
\begin{align}
\mathcal{H}_X = X \sum_{i,\sigma} (n_{i,\bar{\sigma}} - n_{i+1,\bar{\sigma}})^2 B_{i,i+1,\sigma}
\end{align}
with $X>0$ ($X\leq0$), where $\bar\sigma$ is the opposite spin to $\sigma$~\cite{Japaridze1995, Japaridze1997, Japaridze1999, Aligia2000, Dobry2010, Nakamura2000, DiLiberto2014}. 
There are two kinds of transitions, a Gaussian transition between the two gapped states and a spin-gap transition belonging 
to the universality class of the level-1 SU(2) Wess-Zumino-Witten model in the weak coupling region.
For $X\leq0$, the Gaussian transition occurs between the CDW and BCDW phases, and the spin-gap transition occurs 
between the MI and BCDW phases.
For $X>0$, the order of the Gaussian and spin-gap transitions changes, i.e., the Gaussian transition between the MI and BSDW phases, 
and the spin-gap transition between the CDW and BSDW phases. 

Interestingly, such a correlated-hopping interaction appears in the Floquet effective model of the periodically-driven Hubbard model.
It has been suggested that periodic driving can induce interaction-/density-dependent photon-assisted tunneling and exchange interactions~\cite{Sias2008, Oka2009, Chen2011, Simon2011, Ma2011, Daley2014, Bermudez2015, Mentink2015, Bukov, Eckardt2015, Itin2015, Meinert2016, Bukov2016, Kitamura2016, Mikami2016, Agarwala2017, Coulthard2017, Desbuquois2017, Goerg2018, Barbiero2018, Messer2018, Hejazi2018}.
For example, we choose a vector potential $A(t)=A_{0}\cos \Omega t$ describing an electric field $E(t)=V_{0}\sin \Omega t$ with the amplitude of the electric field $V_{0}$ and vector potential $A_{0}=\frac{V_{0}}{\Omega}$.
Setting the frequency $\Omega$ such that $U=l\Omega$ ($\gg t_h$), where $l$ is a natural number, the effective Hamiltonian to the lowest order in $1/\Omega$ is given by
\begin{subequations}
\begin{align}
\mathcal{H}_{\text{res}}^{\text{eff}} 
=& \sum_{i, \sigma} \bigl\{ -J_{\text{eff}}g_{i,i+1,\sigma} -K_{\text{eff}} [h_{i,i+1,\sigma}^\dag + (-1)^{l}h_{i+1,i,\sigma}^\dag] \nonumber \\ 
&+ \text{H.c.} \bigr\} \\
=& \sum_{i, \sigma} \bigl\{ J_{\text{eff}} [-B_{i,i+1,\sigma} + (n_{i,\bar{\sigma}} - n_{i+1,\bar{\sigma}})^2 B_{i,i+1,\sigma}] \nonumber \\
&-K_{\text{eff}}(n_{i,\bar{\sigma}} - n_{i+1,\bar{\sigma}})^p B_{i,i+1,\sigma} \bigr\},
\end{align}
\end{subequations}
where $J_{\text{eff}}=t_h \mathcal{J}_0(A_0)$ and $K_{\text{eff}}=t_h \mathcal{J}_l(A_0)$.
Here, $\mathcal{J}_l$ is the Bessel function of the first kind of order $l$.
If $l$ is odd (even), $p=1$ ($p=2$).
$g_{i,j,\sigma}=(1-n_{i,\bar{\sigma}}) c_{i,\sigma}^\dag c_{j,\sigma} (1-n_{j,\bar{\sigma}}) + n_{i,\bar{\sigma}} c_{i,\sigma}^\dag c_{j,\sigma} n_{j,\bar{\sigma}}$ represents the hopping of doublons and holons, and $h_{i,j,\sigma}^\dag=n_{i,\bar{\sigma}} c_{i,\sigma}^\dag c_{j,\sigma} (1-n_{j,\bar{\sigma}})$ creates a doublon-holon pair.
Since $\mathcal{H}_{\text{res}}^{\text{eff}} $ contains the correlated hopping term similar to $\mathcal{H}_X$, bond-located order can be induced by the driving pulse.
In particular, as long as the intensity of the external field is not extremely strong, $J_{\text{eff}}$ remains positive, which corresponds to $X>0$, and thus the driving pulse may give rise to BSDW.

We demonstrate that BSDW is indeed induced in the 1DEHM driven by a photon pulse.
We assume that the pulse has the time dependence determined by the vector potential $A(t)=A_0 e^{-(t-t_0)^2/(2t_\mathrm{d}^2)} \cos[\Omega(t-t_0)]$.
Spatially homogeneous electric field applied along the chain in the Hamiltonian is incorporated via the Peierls substitution in the hopping terms as $c_{i,\sigma}^\dag c_{i+1,\sigma} \rightarrow e^{iA(t)}c_{i,\sigma}^\dag c_{i+1,\sigma}$~\cite{Peierls1933, Madsen2002}.

\section{Machine learning}\label{sec_3}
Recent extensive studies have demonstrated that ML can provide new strategies for investigating quantum states~\cite{Mehta2018}.
This approach has been successful in characterizing ordered states~\cite{Carrasquilla2017, Broecker2017, Chng2017}, many-body localized states~\cite{Schindler2017, Venderley2018}, topological states~\cite{Ohtsuki2016, Zhang2017, Zhang2017_2, Ohtsuki2017, Broecker2017_2, Nieuwenburg2017, Deng2017, Yoshioka2018, Suchsland2018}, Floquet-engineered states~\cite{Bukov2018}, and even experimental data~\cite{Rem2018, Bohrdt2018}.
We expect that the supervised ML is also useful for examining photoexcited states.

ES 
is useful for characterizing phases of not only 1DEHM~\cite{Gu2004,Deng2006,Anfossi2007,Mund2009,Iemini2015,Yu2016}, but also various other systems~\cite{Li2008, Regnault2009, Yao2010, Lauchli2010, Thomale2010, Turner2010, Fidkowski2010, Thomale2010_2, Lundgren2012, Lepori2013, Giampaolo2013, James2013, Bayat2014, Shinjo2015, Shirakawa2016, Mandal2016}.
We therefore use ES as a training dataset in the present study.
In order to calculate ES, we use the density-matrix renormalization group (DMRG) method.
In a system composed of two subsystems A and B, a Schmidt decomposition of a many-body state $|\psi \rangle$ reads
\begin{align}
|\psi \rangle = \sum _{i=1} {\bar p}_i |\psi^i_A \rangle |\psi ^i _B \rangle =\sum _{i=1} e^{-\xi_i} |\psi _A^i \rangle |\psi _B^i \rangle,
\end{align} 
where ${\bar p}_i$ is the eigenvalue of reduced density matrix $\rho _A ={\rm Tr}_B |\psi \rangle \langle \psi | =e^{-\mathcal{H}_E}$ for subsystem A (or $\rho _B ={\rm Tr}_A |\psi \rangle \langle \psi |$ for subsystem B) and ES $\xi_i$ (in ascending order) is the eigenvalue of the entanglement Hamiltonian $\mathcal{H}_E$.
We take the subsystem A be half of the whole system throughout this paper.

\begin{figure}[t]
  \centering
    \includegraphics[clip, width=20pc]{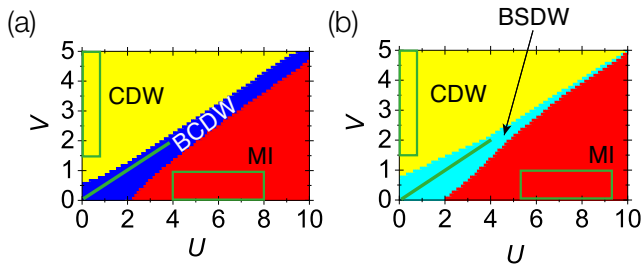}
    \caption{Ground-state phase diagram of the half-filled 1DEHM with $\mathcal{H}_X$ constructed by ML for (a) $X=-1/4$ and (b) $X=1/4$. MI, CDW, BCDW, and BSDW are shown by red, yellow, blue, and light blue colors, respectively. Training dataset is extracted randomly from the regions indicated by green rectangles and lines~\cite{ML_training}.}
    \label{f1}
\end{figure}

We calculate a gap $\Delta\xi_i=\xi_{i+1}-\xi_i$ called the Schmidt gap for the ground state of 1DEHM with the correlated hopping term $\mathcal{H}_X$ for system size $L=20$ under open boundary conditions.
The training dataset is extracted randomly from the regions indicated by green rectangles and lines in Fig.~\ref{f1}~\cite{ML_training}.
The total number of the dataset is 20000.
Using this dataset, we construct a four-layer neural network with two hidden layers, where there are 200 input units for $\Delta\xi_i$ ($i=1,2,\dots,200$), 300 hidden units for each hidden layer, and 4 output units to distinguish the four phases.
Our network is trained and optimized using the ground states only with the help of the Chainer framework~\cite{Chainer} (see Appendix \ref{AA}).
Both training and test errors of the network are found to be less than $0.001\%$.
Using the network, the ground state phase diagram of the 1DEHM with $\mathcal{H}_X$ is obtained as shown in Fig.~\ref{f1}(a) for $X=-1/4$ and Fig.~\ref{f1}(b) for $X=1/4$, which are in good agreement with the phase diagrams obtained previously~\cite{Nakamura2000, Shinjo2019}.
One finds that BCDW is stable for $X\leq0$, while BSDW is stable for $X>0$ in the intermediate phase.
Since the finite size effect is expected to be particularly large in the small $U$ region, we need to compute in larger systems for DMRG calculations to improve accuracy of the boundaries, which remains as a future work.

\begin{figure}[t]
  \centering
    \includegraphics[clip, width=20pc]{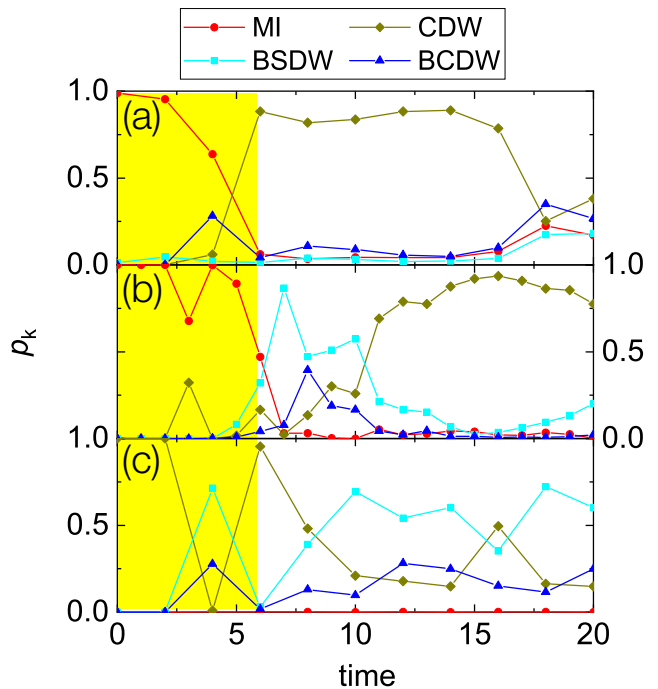}
    \caption{Time evolution of the likelihood $p_k$ with $k=\text{MI, CDW, BCDW, and BSDW}$ in the $L=20$ half-filled 1DEHM for (a) $(U,V,\Omega)=(10,4.5,4)$, (b) $(U,V,\Omega)=(10,1.5,6)$, and (c) $(U,V,\Omega)=(10,7,10)$. Here, $\Omega$ is chosen such that the ground states for each $U$ and $V$ are driven with near-resonant frequency. A photon pulse is applied during the yellow-shaded time region. The likelihoods are obtained by ML.}
    \label{f2}
\end{figure}

We apply driving pulse with $A_0=0.5, t_d=0.5$, and $ t_0=3$ to the 1DEHM for $(U,V)=(10,1.5)$, $(10,4.5)$, and $(10,7)$.
Before applying a pulse, the ground state is MI (CDW) state for $(U,V)=(10,1.5)$ and $(10,4.5)$ [$(U,V)=(10,7)$].
We expect that the photoinduced state is predominantly characterized by some of the phases that are located near the ground state before driving.
The time evolution of ES is calculated for the $L=20$ 1DEHM using the time-dependent DMRG method, which has widely been applied to investigate nonequilibrium phenomena~\cite{White2004,Daley2004,Manmana2005,Haegeman2011,Zaletel2015}. 
Time evolutions of the likelihood $p_k$ ($k$ indicates each phase) are shown in Fig.~\ref{f2} for these three sets of $(U,V)$. 

If the ground states are driven by a photon pulse with near-resonant frequency such as $\Omega=6$, 4, and 10 for $(U,V)=(10,1)$, $(U,V)=(10,4.5)$ and $(U,V)=(10,7)$, respectively (see Appendix \ref{AB}), the likelihood changes with time. 
We find that the photoexcited states predicted by ML are different for different values of $V$.
The case of $(U,V)=(10, 4.5)$ shown in Fig.~\ref{f2}(a) demonstrates that the likelihood obtained by ML correctly predicts photoexcited states, since the photoinduced state driven by a photon pulse with $\Omega=4$ has been suggested to be CDW~\cite{Lu2012}.
Calculating charge correlation functions, we indeed find the enhancement of CDW order in the photoinduced state for $L=20$, which demonstrates that ML has the ability to characterize photoexcited states of the 1DEHM.
For the case of $(U,V)=(10,1.5)$ shown in Fig.~\ref{f2}(b), $p_{\text{BSDW}}$ first increases just after the potoirradiation but soon decreases, followed by strong increase of $p_{\text{CDW}}$, indicating the enhancement of BSDW and CDW orders by photoexcitation.
For the case of $(U,V)=(10,7)$ shown in Fig.~\ref{f2}(c), $p_{\text{BSDW}}$ is the largest, implying that BSDW order is enhanced by photoexcitation.
We also notice that in both cases of $(U,V)=(10,1.5)$ and $(10,7)$, when $p_{\text{BSDW}}$ increases, $p_{\text{CDW}}$ decreases, and vice versa.
The prediction by ML being a clue, we further investigate photoexcited states by explicitly calculating the time evolution of physical quantities focusing on BSDW and CDW.

\section{Photoinduced orders}\label{sec_4}
In order to organize the order parameters to be considered in the 1DEHM, a low-energy effective model obtained by a bosonization treatment is considered~\cite{Giamarchi}.
Upon neglecting terms of higher scaling dimension, the effective Hamiltonian turns out to be the sum of two decoupled sine-Gordon model,
\begin{align}\label{eq:hs}
\mathcal{H} =& \sum_{\nu=c,s} \frac{v_\nu}{2\pi} \int_0 ^L dx \left[ K_\nu (\partial_x \theta_\nu)^2 + K_\nu ^{-1} (\partial_x \phi_\nu)^2 \right] \nonumber \\
&+ \sum_{\nu=c,s} \frac{2m_\nu v_\nu}{(2\pi \alpha)^2} \int _0^L dx \cos [\sqrt{8}\phi_\nu (x)],
\end{align}
where the bosonic fields $\phi_\nu(x)$ and $\theta_\nu(x)$ satisfy the commutation relation $[\phi_\mu (x), \theta_\nu (x')] = -\frac{i\pi}{2} \delta_{\mu \nu} \text{sgn} (x-x')$.
$K_\nu$, $v_\nu$, and $m_\nu$ are the Luttinger parameters, the velocities, and the masses, respectively, for charge 
($\nu=c$) and spin ($\nu=s$) sectors.
An important criterion to classify the various phases is to identify the presence of charge and/or spin gaps.
The opening of a gap takes place whenever the vacuum expectation value $\langle \phi_\nu \rangle$ of the corresponding field pins to a value that minimizes the cosine in Eq.~(\ref{eq:hs}).
As far as the charge sector is concerned, a gap can open only at half-filling, and there are two possible sets of pinning values for $\phi_c$ depending on the sign of $m_c$.
If the spin gap is closed, these pinning values correspond to the two possible insulators, which are denoted as MI and BSDW also known as charge-gapped Haldane insulator (HI).
In contrast, a spin gap always opens when $m_s<0$ due to SU(2) invariance in the spin sector, where only one way of pinning $\phi_s$ is possible. 
In this case, when the charge gap is closed, the spin gapped phase is the Luther-Emery (LE) phase.
When the charge gap is also open, one has two possible fully gapped phases: for $m_c < 0$ the BCDW phase and for $m_c > 0$ the CDW phase, whose order parameters can be written as $\mathcal{O}_{\text{BCDW}}(j) = (-1)^j (B_{j,j+1,\uparrow}+B_{j,j+1,\downarrow})$ $\propto \cos[\sqrt{2}\phi_c(x)] \cos[\sqrt{2}\phi_s(x)]$ and $\mathcal{O}_{\text{CDW}}(j) = (-1)^j n_j$ $\propto \sin[\sqrt{2}\phi_c(x)] \cos[\sqrt{2}\phi_s(x)]$, respectively.

\begin{figure}[t]
  \centering
    \includegraphics[clip, width=20pc]{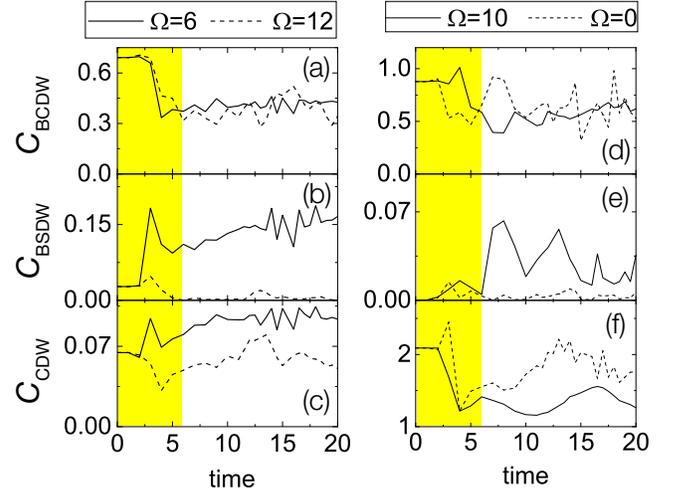}
    \caption{Time evolution of LOPs for [(a) and (d)] BCDW, [(b) and (e)] BSDW, and [(c) and (f)] CDW calculated in the $L=20$ half-filled 1DEHM with $(U,V)=(10,1.5)$ for (a)--(c) and $(U,V)=(10,7)$ for (d)--(f). Correlation functions for near-resonant (off-resonant) driving are shown by solid lines (dashed lines).  A photon pulse is applied during the yellow-shaded time region. These are calculated by the time-dependent DMRG method.} 
    \label{f3}
\end{figure}

We investigate the time evolution of correlation functions of local order parameters (LOPs) $\mathcal{C}_{\text{BCDW}}$, $\mathcal{C}_{\text{BSDW}}$, and $\mathcal{C}_{\text{CDW}}$ for BCDW, BSDW, and CDW orders, respectively, which can be defined as $ \mathcal{C}_\kappa= \frac{1}{\mathcal{N}}\sum_j \mathcal{C}_\kappa (j)$ with $\mathcal{C}_\kappa (j)=\sum_k (-1)^{|j-k|} \langle \mathcal{P}_\kappa (j) \mathcal{P}_\kappa (k) \rangle$, $\kappa = \text{BCDW, BSDW, and CDW}$, and $\mathcal{N}=L$ ($\mathcal{N}=L-1$) for $\kappa=\text{CDW}$ (BCDW and BSDW). 
LOPs are defined as 
\begin{subequations}
\begin{align}
\mathcal{P}_{\text{BCDW}}(j) = B_{j,j+1,\uparrow} + B_{j,j+1,\downarrow},
\end{align}
\begin{align}
\mathcal{P}_{\text{BSDW}}(j) = B_{j,j+1,\uparrow} - B_{j,j+1,\downarrow},
\end{align}
 and 
 \begin{align}
 \mathcal{P}_{\text{CDW}}(j) = n_j-1.
 \end{align}
 \end{subequations}
Time-dependent wave function is calculated by the time-dependent DMRG method under open boundary conditions with keeping 1000 to 2000 density-matrix eigenstates.
Figures~\ref{f3}(a)--\ref{f3}(c) [\ref{f3}(d)--\ref{f3}(f)] show the time evolution of these correlation functions for $(U,V)=(10,1.5)$ 
[$(U,V)=(10,7)$], where the ground state before driving is MI (CDW).
We find that $\mathcal{C}_{\text{BCDW}}$ in Figs.~\ref{f3}(a) and \ref{f3}(d) are suppressed (see Appendix \ref{AC}), while $\mathcal{C}_{\text{BSDW}}$ in Figs.~\ref{f3}(b) and \ref{f3}(e) are enhanced when the near-resonant conditions are satisfied. 
Whether $\mathcal{C}_{\text{CDW}}$ is enhanced or suppressed depends on $V$: $\mathcal{C}_{\text{CDW}}$ for $(U,V)=(10,1.5)$ in Fig.~\ref{f3}(c) is enhanced, while $\mathcal{C}_{\text{CDW}}$ for $(U,V)=(10,7)$ in Fig.~\ref{f3}(f) is suppressed.
Therefore, BSDW in addition to CDW is enhanced for $(U,V)=(10,1.5)$ and only BSDW is enhanced for $(U,V)=(10,7)$ in the photoexcited state. 
These results are in good agreement with the prediction by ML, which indicates that the emergence of BSDW order suggested by the Floquet theory is successfully captured both by ML and explicit calculation of correlation functions.

\begin{figure}[t]
  \centering
    \includegraphics[clip, width=20pc]{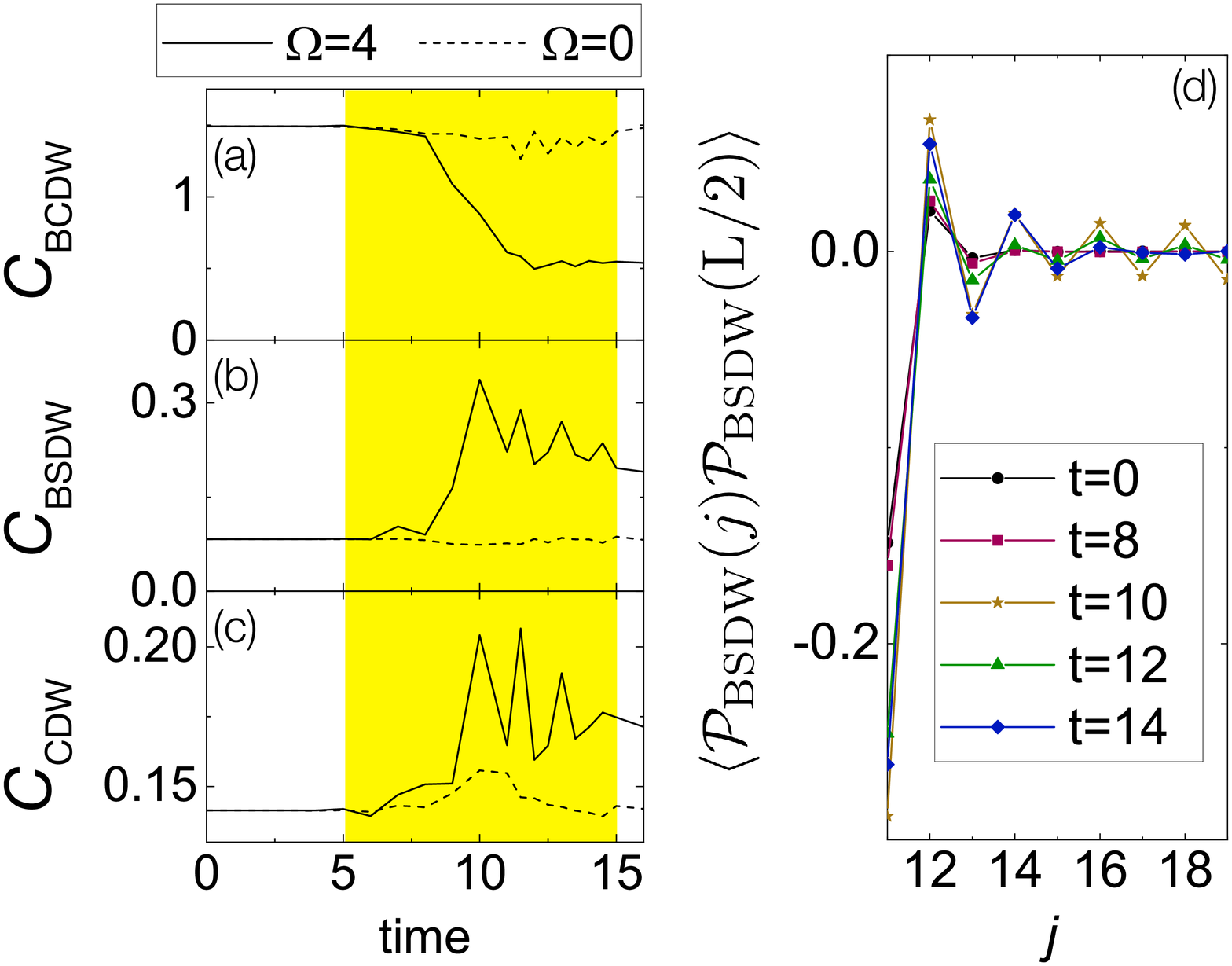}
    \caption{
    Time evolution of (a) BCDW, (b) BSDW, and (c) CDW correlation functions, i.e., $\mathcal{C}_{\text{BCDW}}$,  
    $\mathcal{C}_{\text{BSDW}}$, and $\mathcal{C}_{\text{CDW}}$, respectively, calculated in the $L=20$ half-filled 1DEHM for 
    $(U,V)=(6,1)$. We set $A_0=0.3$, $t_d=2$, and $t_0=10$ with $\Omega=4$ (solid lines) and $\Omega=0$ (dashed lines). 
     A photon pulse is applied during the yellow-shaded time region. 
     (d) $ \langle \mathcal{P}_{\text{BSDW}} (j) \mathcal{P}_{\text{BSDW}} ($L/2$) \rangle$ calculated for time 
     $t=0$ (before driving), 8, 10, 12, and 14. The model parameters are the same as in (a)--(c) except for $\Omega=4$. 
     }
    \label{f4}
\end{figure}

We expect that the photon-assisted correlated-hopping interaction plays a role similar to the $X$ term in the 1DEHM.
In fact, we find that the BSDW correlation function is enhanced in a wide range of $U>0$ and $V>0$ as long as the electric field 
with near-resonant frequency continues to apply to the 1DEHM. 
Figures~\ref{f4}(a)--\ref{f4}(c) shows the time evolution of correlation functions $\mathcal{C}_{\text{BCDW}}$, $\mathcal{C}_{\text{BSDW}}$, 
and $\mathcal{C}_{\text{CDW}}$ for $(U,V)=(6,1)$ with $A_0=0.3$, $t_d=2$, and $t_0=10$. 
Notice that here $t_d=2$ is four time larger than the value set in Fig.~\ref{f3}.
The BSDW and CDW correlations are enhanced in the photoexcited state, while the BCDW correlation is suppressed, 
under the near-resonant condition $\Omega=4$ indicated by the solid lines in Figs.~\ref{f4}(a)--\ref{f4}(c).
The results for $\Omega=0$ indicated by the dashed lines in these figures are examples for the case of off-resonant driving. 
The enhancement of BSDW correlation functions $ \langle \mathcal{P}_{\text{BSDW}} (j) \mathcal{P}_{\text{BSDW}} ($L/2$) \rangle$ in the photoexcited state is focused in Fig.~\ref{f4}(d) for each time $t=0$ 
(before driving), 8, 10, 12, and 14 under the near-resonant condition $\Omega=4$.

\begin{figure}[t]
  \centering
    \includegraphics[clip, width=20pc]{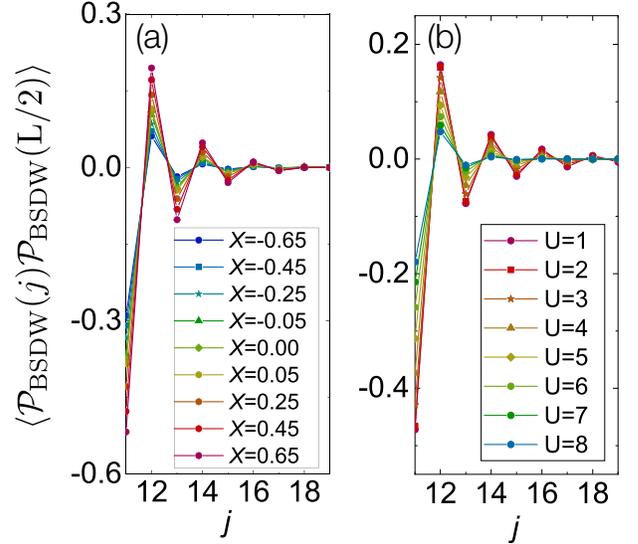}
    \caption{BSDW correlation function $ \langle \mathcal{P}_{\text{BSDW}} (j) \mathcal{P}_{\text{BSDW}} ($L/2$) \rangle$ for the $L=20$ half-filled 1DEHM with the correlated-hopping term $\mathcal{H}_X$. (a) $X$ dependence for $(U,V)=(3,1.5)$. (b) $U$ dependence for $V=U/2$ and $X=0.25$. Note that the ground state for all the parameters studied here except for $X\le 0$ is in the BSDW phase.}
    \label{f5}
\end{figure}

In Fig.~\ref{f5}(a), we show $X$ dependence of the BSDW correlation function $ \langle \mathcal{P}_{\text{BSDW}} (j) \mathcal{P}_{\text{BSDW}} ($L/2$) \rangle$ of the ground state of $\mathcal{H}+\mathcal{H}_{X}$ for $(U,V)=(3,1.5)$.
As expected, the BSDW correlation becomes larger with increasing $X$, revealing the role of correlated-hopping interaction 
$\mathcal{H}_X$ in the 1DEHM.
Figure~\ref{f5}(b) shows the change in the magnitude of BSDW correlation function when $U$ is varied for $V=U/2$ and $X=0.25$.
These results in Fig.~\ref{f5} provide a measure of the magnitude of the BSDW correlation function when the ground state is known 
to be in the BSDW phase.
Comparing Fig.~\ref{f4}(d) with Fig.~\ref{f5}, indeed the irradiation in the 1DEHM plays a role similar to introducing 
the $X$ term in the 1DEHM: The photoirradiation temporally generates the Floquet state with the enhanced BSDW correlations. 
Notice also in Figs.~\ref{f4}(a)--\ref{f4}(c) that the enhancement of BSDW and CDW correlations in the photoexcited state 
remains even after the pulse is decayed. 

\begin{figure}[t]
  \centering
    \includegraphics[clip, width=20pc]{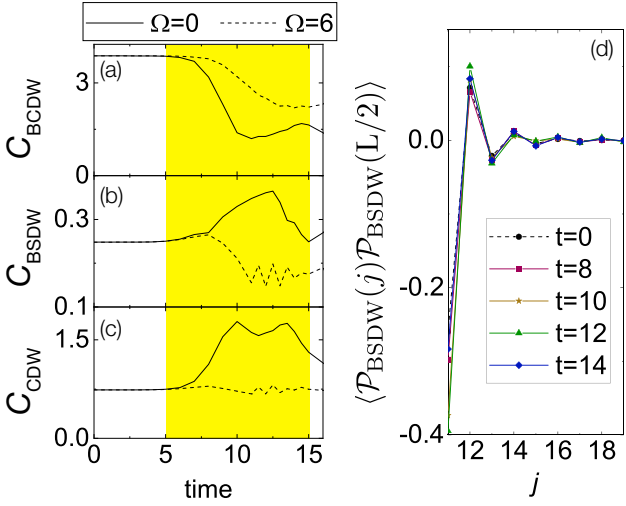}
    \caption{
    Same as Fig.~\ref{f4} but $(U,V)=(6,3)$, $\Omega=0$ (solid lines) and $\Omega=6$ (dashed lines) in (a)--(c), and  
    $\Omega=0$ in (d). 
    }
    \label{f6}
\end{figure}

Figures~\ref{f6}(a)--\ref{f6}(c) show the time evolution of correlation functions $\mathcal{C}_{\text{BCDW}}$, $\mathcal{C}_{\text{BSDW}}$, and $\mathcal{C}_{\text{CDW}}$ for $(U,V)=(6,3)$ with $A_0=0.3$, $t_d=2$, and $t_0=10$. 
As in the case for $(U,V)=(6,1)$, the BSDW and CDW correlations are enhanced in the photoexcited state, while the BCDW correlation 
is suppressed under near-resonant condition. 
The enhancement of the BSDW correlation functions  in the photoexcited state is focused in Fig.~\ref{f6}(d) for each time 
$t=0$ (before driving), 8, 10, 12, and 14 under the near-resonant condition $\Omega=0$. 
Notice that in contrast to the case for $(U,V)=(6,1)$, the enhancement of BSDW correlations seems limited only during the light irradiation 
indicated by the yellow-shaded time region in Fig.~\ref{f6}(b).

We now consider nonlocal order parameters (NLOPs) to examine the mechanism of photoinduced suppression and enhancement in more detail.
Recently it has been shown that NLOPs give an accurate description of bond-order waves~\cite{Montorsi2012, Barbiero2013}.
In addition, NLOPs can detect the presence of non-trivial topological phases~\cite{Montorsi2017, Fazzini2017} and hidden orders not captured by conventional order parameters~\cite{DellaTorre2006, Berg2008, denNijs1989, Zaanen2001, Kruis2004}.
Furthermore, we can directly compare the calculated NLOPs with experimental observation in optical lattices~\cite{Endres2011, Endres2013, Hilker2017}.
Thus, we expect that NLOPs are also useful for characterizing photoexcited bond-order waves.

Let us introduce the parity (P) and string (S) non-local operators~\cite{denNijs1989, Montorsi2012, Barbiero2013}: 
\begin{subequations}
\begin{align}
\mathcal{P}_P^\nu(j) =\prod_{k=0}^{j-1} e^{i\pi S_{z,k}^\nu}
\end{align}
and
\begin{align}
\mathcal{P}_S^\nu(j) =\left( \prod_{k=0}^{j-1} e^{i\pi S_{z,k}^\nu} \right) S_{z,j}^\nu
\end{align}
\end{subequations}
with charge operator $S_{z,j}^c=n_j-1$ and spin operator $S_{z,j}^s=n_{j,\uparrow}-n_{j,\downarrow}$.
Their correlation functions $\mathcal{C}_P^\nu (r)= \langle \mathcal{P}_P^\nu(j)^\dag \mathcal{P}_P^\nu(j+r) \rangle =  \langle e^{i\pi \sum_{l=j}^{l=j+r}S_l^{(\nu)}} \rangle$ and $\mathcal{C}_S^\nu (r) =  \langle \mathcal{P}_S^\nu(j)^\dag \mathcal{P}_S^\nu(j+r) \rangle = \langle S_j^{(\nu)} e^{i\pi \sum_{l=j+1}^{l=j+r-1}S_l^{(\nu)}} S_{j+r}^{(\nu)}\rangle$ remain finite in the limit $r \rightarrow \infty$, 
which corresponds to the opening of a specific gap in the $\nu$ sector~\cite{Barbiero2013}. 
Here we set $r = L/2$.
In particular, within the bosonization approximation, $\mathcal{C}_P^\nu (x) = \langle \cos [ \sqrt{2} \phi_\nu (0)] \cos [ \sqrt{2} \phi_\nu (x)] \rangle$ remains finite when $\phi_\nu (x)$ pins to 0, whereas $\mathcal{C}_S^\nu (x) = \langle \sin [ \sqrt{2} \phi_\nu (0)] \sin [ \sqrt{2} \phi_\nu (x)] \rangle$ is finite when $\phi_\nu (x)$ pins to $\frac{\pi}{\sqrt{8}}$.
Thus, the expectation value of $\mathcal{C}_q^\nu (x)$ ($q=$ P and S) configures as order parameter for the different gapped phases, and is useful for the classification of 1D quantum phases. 
As shown in Table~\ref{table}, there are Luttinger liquid (LL), three partly gapped states such as MI, LE, and BSDW/HI states, and two fully gapped states, i.e., CDW and BCDW states.
These states are characterized by the NLOPs.

\begin{table}[b]
\centering
	\caption{Classification of 1D quantum phases and the corresponding nonlocal order parameters (NLOPs) from bosonization and renormalization group analysis~\cite{Montorsi2012, Barbiero2013, Montorsi2017, Fazzini2017}. $\Delta_c$ and $\Delta_s$ are charge and spin gaps, respectively. Unlocked fields are indicated by $u$.}
  \begin{tabular}{|l||c|c|c|c|c|} \hline
  	   & $\phi_c$ & $\phi_s$ & $\Delta_c$ & $\Delta_s$ & NLOP \\ \hline \hline
    LL & $u$ & $u$ & 0 & 0 & none \\ 
    MI & 0 & $u$ & $\neq 0$ & 0 & $\mathcal{C}_P^{(c)}$ \\
    LE & $u$ & 0 & 0 & $\neq 0$ & $\mathcal{C}_P^{(s)}$ \\
    BSDW/HI & $\pi/\sqrt{8}$ & $u$ & $\neq 0$ & 0 & $\mathcal{C}_S^{(c)}$ \\
    CDW & $\pi/\sqrt{8}$ & 0 & $\neq 0$ & $\neq 0$ & $\mathcal{C}_S^{(c)}$, $\mathcal{C}_P^{(s)}$ \\
    BCDW & 0 & 0 & $\neq 0$ & $\neq 0$ & $\mathcal{C}_P^{(c)}$, $\mathcal{C}_P^{(s)}$ \\ \hline
  \end{tabular}
  \label{table}
\end{table}

\begin{figure}[t]
  \centering
    \includegraphics[clip, width=20pc]{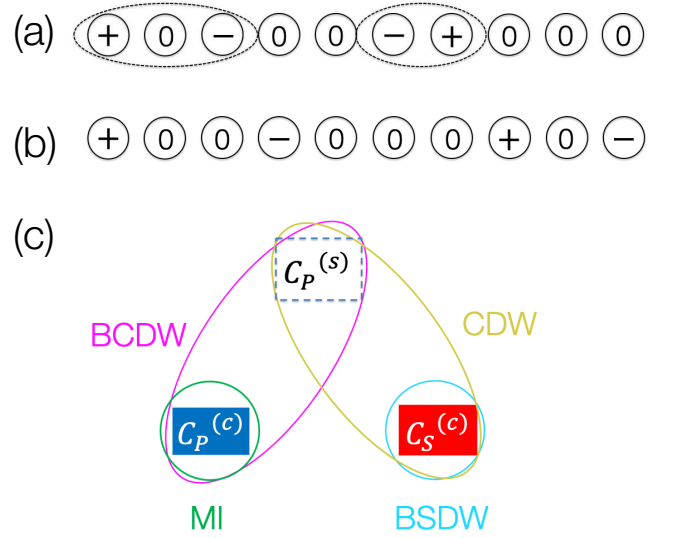}
    \caption{
    Typical configurations of charge fluctuations in (a) the MI state and (b) the BSDW/HI state. 
    ``$+$", ``0", and ``$-$" represent the deviation of the local occupation from the averaged density. 
    The charge fluctuation present in the MI state is described as bound particle-hole pairs indicated by the dotted lines. 
    The charge fluctuation in the BSDW/HI state is characterized as a charge order $+,-,+,-\dots$ with the undetermined number of 0 
    sites between each $+$ and $-$. (c) The mutual relation between NLOPs and LOPs in photoexcited states, which partly visualizes 
    Table~\ref{table}. $\mathcal{C}_S^{(c)}$ enhanced in phtoexcited states is emphasized by red shade, while $\mathcal{C}_P^{(c)}$ 
    suppressed in phtoexcited states is indicated by blue shade.}
    \label{fs-NLOPs}
\end{figure}

In addition, the NLOPs can give crucial informations regarding the microscopic structure encoded in the ground state. 
Indeed, a finite value of the parity in the charge (spin) sector implies that correlated holon-doublon (up-down spin) virtual 
excitations are present. 
On the other hand, a non-zero charge (spin) string order indicates the presence of dilute holon-doublon (up-down spin) staggered order~\cite{Montorsi2012, Barbiero2013}.
Such microscopic information in charge sector is schematically shown in Fig.~\ref{fs-NLOPs}(a) for the parity correlation in the 
MI state and in Fig.~\ref{fs-NLOPs}(b) for the string correlation in the BSDW/HI state.
Here, ``$+$", ``0", and ``$-$" represent the deviation of the local occupation from the averaged density.
The charge fluctuation present in the MI state is described as bound particle-hole pairs indicated by the dotted lines, 
and the charge fluctuation in the BSDW/HI state is characterized as a charge order $+,-,+,-\dots$ 
with the undetermined number of 0 sites between each $+$ and $-$.

As we will see, $\mathcal{C}_S^{(c)}$ ($\mathcal{C}_P^{(c)}$) is always enhanced (suppressed) by 
photoexcitation under near-resonant conditions. 
If the MI ground state of 1DEHM is photoexcited, holon and doublon are delocalized, i.e., 
$\mathcal{C}_P^{(c)}$ is suppressed.
Simultaneously, the holon-doublon fluctuation with holons and doublons appearing alternatingly along the chain is enhanced, 
and thus $\mathcal{C}_S^{(c)}$ is enhanced.
Whether $\mathcal{C}_P^{(s)}$ is enhanced or suppressed depends on $V$.
The mutual relation between NLOPs and LOPs is summarized schematically in Fig.~\ref{fs-NLOPs}(c).
This relation implies that MI (BSDW) is not enhanced (suppressed) but suppressed (enhanced) in photoexcited states. 
When $\mathcal{C}_P^{(s)}$ is enhanced (suppressed), CDW is expected to be effectively enhanced (suppressed) in photoexcited states.

\begin{figure}[t]
  \centering
    \includegraphics[clip, width=20pc]{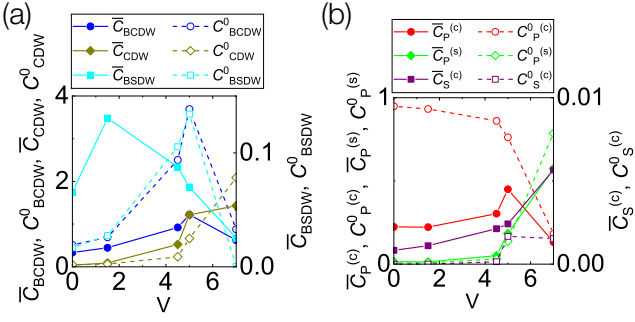}
    \caption{The change of various correlation functions before and after applying the driving pulse calculated in the $L=20$ half-filled 1DEHM with $U=10$. (a) $V$ dependence of $\bar{\mathcal{C}}_{\text{BCDW}}$ ($\mathcal{C}^0_{\text{BCDW}}$), $\bar{\mathcal{C}}_{\text{BSDW}}$ ($\mathcal{C}^0_{\text{BSDW}}$), and $\bar{\mathcal{C}}_{\text{CDW}}$ ($\mathcal{C}^0_{\text{CDW}}$) indicated by solid (open) symbols. (b) $V$ dependence of $\bar{\mathcal{C}}_P^{(c)}$ ($\mathcal{C}_P^{0(c)}$), $\bar{\mathcal{C}}_P^{(s)}$ ($\mathcal{C}_P^{0(s)}$), and $\bar{\mathcal{C}}_S^{(c)}$ ($\mathcal{C}_S^{0(c)}$) indicated by solid (open) symbols. Here we set $\Omega=8$ for $V=0$, $\Omega=6$ for $V=1.5$, $\Omega=4.5$ for $V=4$, $\Omega=2$ for $V=5$, and $\Omega=10$ for $V=7$. The results are calculated by the time-dependent DMRG method.
    }
    \label{f8}
\end{figure}

We show the $V$ dependence of LOPs (NLOPs) in Fig.~\ref{f8}(a) [Fig.~\ref{f8}(b)], where $\bar{\mathcal{{C}}}_\kappa$ ($\bar{\mathcal{{C}}}_q^{(\nu)}$) is the time-averaged value of $\mathcal{C}_\kappa$ ($\mathcal{C}_q^{(\nu)}$) over time $0<t<20$, and $\mathcal{C}_\kappa^0$ ($\mathcal{C}_q^{0(\nu)}$) is the correlation function in the ground state before driving 
with $\kappa=\text{BCDW, BSDW, and CDW}$ ($q=\text{P and S}$, and $\nu=c \text{ and } s$).
Here, we choose frequency $\Omega$ that nearly resonates with the gap: $\Omega=8$, 6, 4, 2, and 10 for $(U,V)=(10,0)$, $(10,1.5)$, $(10,4.5)$, $(10,5)$, and $(10,7)$, respectively.
In these cases, the change of the correlation functions by photoexcitation is the largest.
$\mathcal{C}_S^{(c)}$ ($\mathcal{C}_P^{(c)}$) is always enhanced (suppressed) by photoexcitation under near-resonant conditions, as  $\delta \mathcal{C}_S^{(c)} = \bar{\mathcal{C}}_S^{(c)} - \mathcal{C}_S^{0(c)} > 0$ ($\delta \mathcal{C}_P^{(c)} = \bar{\mathcal{C}}_P^{(c)} - \mathcal{C}_P^{0(c)} < 0$) for $0\leq V\leq 7$, as shown in Fig.~\ref{f8}(b).

Whether $\mathcal{C}_P^{(s)}$ is enhanced or suppressed depends on $V$. 
As shown in Fig.~\ref{f8}(b), $\delta \mathcal{C}_P^{(s)} = \bar{\mathcal{C}}_P^{(s)} - \mathcal{C}_P^{0(s)} > 0$ $(\delta \mathcal{C}_P^{(s)} < 0)$ for $0\leq V \leq 5$ ($5< V \leq 7$), indicating the enhancement (suppression) of CDW. 
In fact, as shown in Fig.~\ref{f8}(a), we find that $\delta \mathcal{C}_{\text{CDW}} = \bar{\mathcal{C}}_{\text{CDW}} - \mathcal{C}_{\text{CDW}}^{0} >0$ for $0\leq V \leq 5$ and $\delta \mathcal{C}_{\text{CDW}}<0$ for $5< V \leq 7$.
The photoinduced enhancement of CDW predicted by ML in Fig.~\ref{f2}(a) is thus confirmed by the direct calculation of the order parameters, which is also in good accordance with the previous study~\cite{Lu2012}.

When CDW is enhanced in photoexcited state, both $\mathcal{C}_P^{(s)}$ and $\mathcal{C}_S^{(c)}$ are simultaneously enhanced. 
Since the BSDW state is related to $\mathcal{C}_S^{(c)}$, BSDW can also be enhanced when CDW is enhanced. 
Indeed, we find that this is the case, which is also predicted by ML in Fig.~\ref{f2}(b). 
As shown in Figs.~\ref{f3}(b) and \ref{f3}(c), after applying the driving pulse, BSDW as well as CDW is enhanced 
and the time-averaged value of $\mathcal{C}_{\text{BSDW}}$ remains finite. 
We find in Fig.~\ref{f8}(a) that $\delta \mathcal{C}_{\text{BSDW}} = \bar{\mathcal{C}}_{\text{BSDW}} - \mathcal{C}_{\text{BSDW}}^0 > 0$ for $V=0$ and 1.5. 
However, as shown in Fig.~\ref{f8} (a), $\delta \mathcal{C}_{\text{BSDW}} <0$ for $V=4.5$ and 5, where the ground state is located near the phase boundary. 
Even in such cases close to the phase boundary, we still find that $\delta \mathcal{C}_{\text{BSDW}} > 0$ as long as the photon pulse with near-resonant frequency continues to apply (see Fig.~\ref{f6}). 
The enhancement of BSDW is also found for $V=7$, as shown in Figs.~\ref{f3}(e) and \ref{f8}(a).
In this case, CDW is suppressed [see Figs.~\ref{f3}(f) and \ref{f8}(a)]. 
This is consistent with $\delta \mathcal{C}_P^{(s)}<0$ and $\delta \mathcal{C}_S^{(c)}  >0$ for $V=7$ [see Fig.~\ref{f8}(b)] and the prediction by ML [see Fig.~\ref{f2}(c)].

\section{Summary}\label{sec_5}
We have proposed and demonstrated that the supervised ML can give useful information for 
characterizing photoexcited states. 
Using ES as a training dataset, the neural network is trained to find the quantum state driven by a photon pulse.
In the cases of $(U,V)=(10,1.5)$ and $(10,7)$, the ML predicts the enhancement of BSDW in photoexcited states.
We have found that the correlation functions of BSDW and the corresponding string NLOP are indeed enhanced by photoexcitation, confirming the prediction by ML.
Combining the prediction by ML and explicit calculation of the correlation functions, we have successfully captured the emergence of BSDW order suggested by the Floquet theory in 1DEHM.
We should emphasize that the enhancement of BSDW found in our study is interesting not only because BSDW is hidden in the ground state phase diagram of the 1DEHM but also has been discussed in the context of the nontrivial symmetry-protected topological states in the continuum limit~\cite{denNijs1989, XChen2011, Schuch2011, Else2013, Montorsi2017, Fazzini2017}. 

Experimentally, it is interesting to observe photoexcited states by pump-probe experiment in tetracyanoquinodimethane (TCNQ) salts and halogen-bridged transition metal compounds, since optical properties in these materials are well-described by the 1DEHM~\cite{Kumar2011, Yamashita1999}.
Since a CDW to MI transition has already been observed~\cite{Kimura2009}, we expect that the enhancement of BSDW in photoexcited states is also detectable in pump-probe spectroscopy.

\begin{acknowledgments}
We thank C. Shao, H. Lu, K. Tanaka, and S. Ejima for useful discussions. This work was supported by CREST (Grant No. JPMJCR1661), the Japan Science and Technology Agency, the creation of new functional devices and high-performance materials to support next-generation industries (CDMSI), and the challenge of basic science exploring extremes through multi-physics and multi-scale simulations to be tackled by using a post-K computer. This work was also supported by a Grants-in-Aids for Young Scientists (B) (No. 17K14148) from MEXT, Japan. The numerical calculation was partly carried out at the facilities of the Supercomputer Center, the Institute for Solid State Physics, the University of Tokyo.
\end{acknowledgments}

\appendix

\section{Machine learning}\label{AA}

\begin{figure}[t]
  \centering
    \includegraphics[clip, width=20pc]{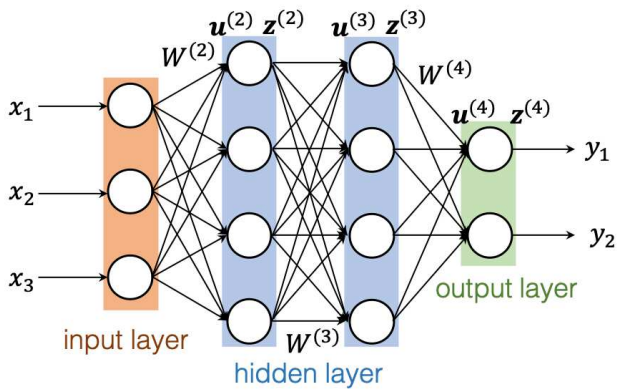}
    \caption{The construction of the neural network is schematically shown. The network consists of input, hidden, and output layers. A input dataset $\bm{x}$ is classified as $\bm{y}$.}
    \label{NL}
\end{figure}

Figure~\ref{NL} shows the neural network constructed in this study.
The initial vector of the input layer is $\bm{z}^{(n=1)}=\bm{x}$, where a training dataset is substituted in input vector $\bm{x}$.
Output vector $\bm{z}^{(n+1)}$ of the $(n+1)$-th layer is written as
\begin{subequations}
\begin{align}
\bm{z}^{(n+1)}=&f(\bm{u}^{(n+1)}), \\
\bm{u}^{(n+1)}=& W^{(n+1)}\bm{z}^{(n)}+\bm{b}^{(n+1)},
\end{align}
\end{subequations}
where $f(u_k^{(n)})$ is the activation function, $W^{(n+1)}$ is the weight matrix, and $\bm{b}^{(n+1)}$ is the bias vector.
$f(u_k^{(n)})=\max(u_k^{(n)},0)$ for the input and hidden layers, while $f(u_k^{(N)})=\frac{\exp \left[u_k^{(N)}\right]}{\sum _{k=1}^4 \exp \left[u_k^{(N)}\right]}$ for the output layer.
Finally, we obtain output vector $\bm{y}=\bm{z}^{(N)}$, where $N$ is the total number of layers.
The output $y_k$ is regarded as the probability $p_k$ that the input dataset belongs to class $k$.
Minimizing the loss function that is defined by comparing output label $y_k$ and input label $d_k$, $W^{(n)}$ and $\bm{b}^{(n)}$ are optimized.
In our case, the number of units in the hidden layers is $3/2$ of that in the input layer, and $N=4$.

In order to increase the accuracy of the prediction by the neural network and also obtain better convergent results, 
we do not use entanglement spectrum $\xi_i$ itself, 
but a Schmidt gap $\xi_{i+1}-\xi_i$, for the input dataset. 
This is because the Schmidt gap can capture more appropriate features of the change of quantum states.
In addition, by using the Schmidt gap, the dataset is appropriately normalized and standardized as often done in data 
preprocessing before machine learning. 
Training and test errors of the network are less than $0.001\%$.

\section{Optical conductivity}\label{AB}
\begin{figure}[t]
  \centering
    \includegraphics[clip, width=20pc]{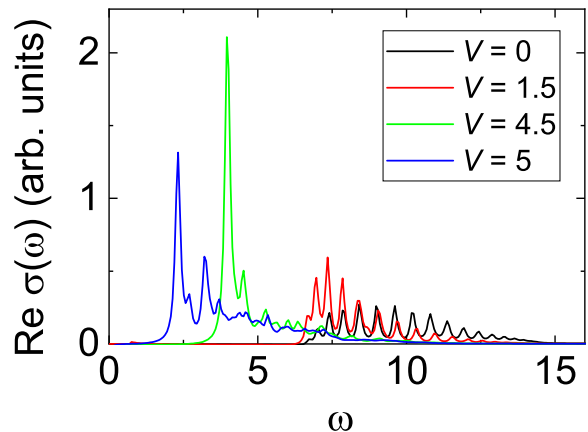}
    \caption{$\text{Re}\sigma (\omega)$ in the half-fillded 1DEHM for $L=20$ with $U=10$ and various values of $V$.}
    \label{OC_L20o}
\end{figure}

Using the method described in Ref~\cite{Lu2015, Shao2016, Shinjo2018}, we obtain optical conductivity $\text{Re}\sigma (\omega)$ of the ground state of the half-filled one-dimensional extended Hubbard model (1DEHM) with system size $L=20$ and open boundary conditions for $(U,V)=(10,0)$, $(10,1.5)$, $(10,4.5)$, and $(10,5)$. 
The result is shown in Fig.~\ref{OC_L20o}.
We obtain $\text{Re}\sigma (\omega)$ by calculating time evolution of the charge current.
We find exciton peaks at $\omega=3.9$ for $(U,V)=(10,4.5)$ and $\omega=2.3$ for $(U,V)=(10,5)$.
For $(U,V)=(10,0)$ and $(10,1.5)$, $\text{Re}\sigma (\omega)$ does not show such exciton peaks, but has the largest weight at $\omega=8.4$ and $\omega=7.3$, respectively.
This explains that the change of correlation functions are the largest for photoexcitation with near-resonant frequency $\Omega$ as discussed in the main text.

\section{BCDW in photoexcited states}\label{AC}
In the parameter region studied in the main text, BCDW is not enhanced, as shown in Fig.~4(a).
However, the enhancement of BCDW has been reported in Ref.~\cite{Shao2018} for $(U,V)=(10,7)$ calculated by the Lanczos 
method under periodic boundary conditions.
We infer that the enhancement of BCDW strongly depends on boundary conditions~\cite{Shao_private}.
Indeed, our calculations under periodic boundary conditions find that $\mathcal{C}_{\text{BCDW}}$ is enhanced in photoexcited states for $L=12$ and $(U,V)=(10,7)$.
In this case, $\mathcal{C}_P^{(c)}$, $\mathcal{C}_S^{(c)}$, and $\mathcal{C}_{\text{BSDW}}$ are also enhanced by photoexcitation.
Further investigation is required to understand why a photoexcited state for the CDW phase strongly depends on boundary conditions, which is left for further study.


\begin{thebibliography}{99}
\bibitem{ML_training} We take the following parameter spaces as training dataset: $\{4<U<8,0<V<1,X=\pm1/4\}$ for MI phase, $\{0<U<1,1.5<V<5,X=\pm1/4\}$ for CDW phase, $\{0<U<4,V=U/2,X=-1/4\}$ for BCDW phase, and $\{0<U<4,V=U/2,X=1/4\}$ for BSDW phase
\end{thebibliography}
\end{document}